\documentstyle[aps]{revtex}
\begin{document}

\title{
Phase diagram of 2D array of mesoscopic granules.
}

\author{A.I. Belousov, S.A. Verzakov and Yu.E. Lozovik\cite{q}}

\address{Institute of Spectroscopy, Russian  Academy  of  Sciences, \\
142092, Troitsk, Moscow region, Russia}

\maketitle

\begin{abstract}
A lattice boson model is used to study ordering phenomena in
regular 2D array of superconductive mesoscopic granules,
Josephson junctions or pores filled with a superfluid helium.
Phase diagram of the system, when quantum fluctuations of both
the phase and local superfluid density are essential,
is analyzed both analytically and by quantum Monte Carlo technique.
For the system of strongly interacting bosons it is found that
as the boson density $n_0$ is increased the boundary
of ordered superconducting state shifts to {\it lower temperatures}
and at $n_0 > 8$ approaches its limiting position
corresponding to negligible relative fluctuations of moduli of the
order parameter (as in an array of "macroscopic" granules).
In the region of weak quantum fluctuations of phases mesoscopic phenomena
manifest themselves up to $n_0 \sim 10$.
The mean field theory and functional integral $1/n_0$ - expansion results
are shown to agree with that of quantum Monte Carlo calculations
of the boson Hubbard model and its quasiclassical limit,
the quantum XY model.
\end{abstract}

\section{Introduction}
\label{introd}
The study of mesoscopic systems has resulted thus far in many
new interesting fundamental concepts \cite{Imry,Aronov}.
Progress has been especially rapid due to
the development of nanotechnology methods which
has opened up new avenues for sophisticated experiments. In this connection
the study of arrays of ultarsmall
granules, microclusters or Josephson junctions is of particular concern
(see e.g. \cite{Zant,Paalanen}).

Granular superconductors, Josephson arrays, superfluid
helium in a porous media \cite{Blum} are,
as a rule, described in terms of different modifications
of quantum XY model (see below), but this description is correct
only if relative fluctuations of the local superfluid density
are not essential \cite{Bruder}. It takes place in sufficiently large
granules at temperatures far below that of onset of superconductivity
in each individual granule. To study the role of quantum fluctuations of
moduli other more adequate models should be used.

A convenient starting point for the description of the $N \times N$
system of interacting bosons (Cooper pairs in granules,
$He$ atoms in pores etc.) is the Bose - Hubbard Hamiltonian:
\begin{eqnarray}
\hat H_{\bf h} =
\frac{t}{2} \sum\limits_{<i,j>}
\left\{ 2 a{^\dagger}_i a_i -  a{^\dagger}_i a_j -
a{^\dagger}_j a_i \right\} +
\frac{U}{2} \sum\limits_{i}
\left\{ a{^\dagger}_i a_i - n_0 \right\}^2
\label{hamiltonian}
\end{eqnarray}
where $a^{\dagger}_i$ ($a_i$) is a boson creation (annihilation)
operator at a site $i = \overline{1,N^2}$;
$t$ is the strength of the hopping between nearest neighbor sites $<i,j>$
and $U>0$ is an on - site repulsive interaction.

The system (\ref{hamiltonian}) has a rich phase diagram \cite{Cha,Trivedi},
containing a Mott insulating phase (at zero temperature),
the superfluid and normal (metal) phases. At a commensurate density
$n_0 = <a^{\dagger}_i a_i>$ (number of bosons is an integer multiple of the
number of sites) and $T=0$ the boson Hubbard model lies in the same
universality class (see \cite{Cha} - \cite{Fisher})
as the quantum XY model with Hamiltonian:
\begin{eqnarray}
\hat H_{\bf xy} =
J \sum\limits_{<i,j>}
\left\{ 1 - \cos{(\varphi_i - \varphi_j)} \right\} -
\frac{U}{2} \sum\limits_i
\left\{ \partial / \partial \varphi_i \right\}^2
\label{hamiltonian_xy}
\end{eqnarray}
where $\varphi_i \in [0,2\pi)$ are phases of the order parameter.
Obviously, at $T \ne 0$ the requirements the density to be commensurate
$n_0 = k$ is too stringent. In the latter case, the behaviour of the system
(\ref{hamiltonian}) will depends {\it continuously} on $n_0$, and
the {\it critical} properties will be {\it the same} in some band
$n_0 = k \pm \delta n_0$, the width $2 \delta n_0$ should decrease as the
temperature is lowered.

The properties of the system (\ref{hamiltonian_xy}),
that have at finite temperatures the superfluid and metallic phases
are described by two dimensionless parameters:
the temperature in units of Josephson coupling constant $T = k_b T / J$
and quantum parameter $q = \sqrt{ U / J}$ which is responsible for
the strength of zero - point fluctuations of phase. Corresponding parameters
of the Hubbard model are $T = k_{\bf b} T / (t n_0)$ and
$q = \sqrt{U / (t n_0)}$.

The general purpose of this communication is comparing of phase diagrams of
models (\ref{hamiltonian}) and (\ref{hamiltonian_xy}) to estimate
the importance of the mesoscopic phenomena in regular 2D systems.
Of prime interest to us is the case of {\it finite temperatures},
apart of intriguing quantum phase transitions which take place
at $T=0$ (see e.g \cite{Otterlo} and references therein).
Sections \ref{theor_mfa} and \ref{theor_exp} present the mean field and
functional integral $1/n_0$ - expansion approaches. From
{\it ab initio} quantum Monte Carlo calculations of different
characteristic quantities (Section \ref{MC}) we determine the
phase diagram $T_{\bf h}^c(q;n_0)$ of the boson Hubbard model at different
densities $n_0$ and compare it (Section \ref{discuss})
with the phase diagram of 2D quantum XY model.

\section{Mean field approximation.}
\label{theor_mfa}
A qualitative estimation of the phase diagram of the boson model
(\ref{hamiltonian}) can be obtained in a simple mean field approximation
(see e.g. \cite{Simanek},\cite{Zimanyi,JETP} and references therein).
The boundary $T^c(q;n_0)$ of ordered state can be found in MFA
from the equation:
\begin{eqnarray}
\label{mf_bound_hubb}
\frac{q^2}{z} = \frac{
\sum\limits_{n=-n_0}^{\infty} \{n+n_0+1\} \left\{
e^{-q^2[n-\eta]^2 / [2T]} - e^{-q^2 [n+1-\eta]^2 / [2T]}
\right\} / \{ 2n + 1 - 2 \eta\}
}
{
n_0 \sum\limits_{n=-n_0}^{\infty} e^{-q^2 \{ n-\eta \}^2 / \{2T\}}
}
\end{eqnarray}
where $\eta = \mu / U - z / (2 q^2 n_0)$, $z$ is the number
of nearest neigbours ($z=4$ for 2D square lattice).
The Equation (\ref{mf_bound_hubb}) on the boundary of ordered state
differs in some details of that of the Reference \cite{Zimanyi}.
Both of them became equivalent in the limit of large densities $n_0$,
with the Equation (\ref{mf_bound_hubb}) more accurate at $n_0 \sim 1$.

The condition on the chemical potential $\mu$ is that the mean number of
particles be equal to $n_0$:
\begin{eqnarray}
\sum\limits_{n=-n_0}^{\infty} n \exp{( -q^2\{ n - \eta \}^2 / \{2T\}) } = 0
\label{mu_n}
\end{eqnarray}
It should be pointed out, that in the limit $n_0 \to \infty$
Equation (\ref{mu_n}) gives $\mu = 0$ and the boundary of ordered state
transforms to that of quantum XY model \cite{Simanek}.

The lines $T^c_{\bf h}(q;n_0)$ of the boson Hubbard model
obtained from Equations (\ref{mf_bound_hubb})
and (\ref{mu_n}) are shown in Figure~1 for different densities $n_0$.
Calculation shows that the lines $T^c_{\bf h}(q;n_0)$
reaches their limiting position, the phase boundary
$T^c_{\bf xy}(q)$ of quantum XY model at $n_0 > 25$.

\section{$1/n_0$ - expansion.}
\label{theor_exp}

To improve the qualitative mean field estimation of the difference
of phase diagrams of models (\ref{hamiltonian}) and (\ref{hamiltonian_xy})
let us represent the partition function
of the Hamiltonian (\ref{hamiltonian}) in a path integral representation
as a trace over a complex c- number Bose field $\Phi$ \cite{Fisher}:
\begin{eqnarray}
\label{Z_1}
Z_{h} = tr\left\{ e^{-S} \right\} =
\int D(\Phi,\Phi^*) e^{-S(\Phi,\Phi^*)},
\\
S(\Phi,\Phi^*) = \int\limits_{0}^{\beta} \left\{
\i \sum\limits_i \dot \Phi_i \Phi_i^* +
\frac{t}{2} \sum\limits_{<i,j>} |\Phi_i - \Phi_j|^2  +
\frac{U}{2} \sum\limits_i \left[ |\Phi_i|^2 - n_0 \right]^2
\right\} d \tau,
\nonumber
\\
\Phi_i(0) = \Phi_i(\beta),\qquad \Phi_i^*(0) = \Phi_i^*(\beta)
\nonumber
\end{eqnarray}
The substitution
$\Phi_i = \sqrt{n_0 + \delta n_i} e^{\i \varphi_i}$
(at integer $n_0$) in Equation (\ref{Z_1}) gives:
\begin{eqnarray}
\label{Z_2}
Z_{\bf h} = \int D(\delta n,\varphi) e^{-S(\delta n,\varphi)},
\\
S(\delta n,\varphi) = \int\limits_{0}^{\beta} \left\{
\frac{U}{2}\sum\limits_{i} [ \delta n_i ]^2 +
\i \sum\limits_i \delta n_i \dot \varphi_i
\right.
\nonumber
\\
\left.
+ t n_0 \sum\limits_{<i,j>} \left[ 1 +
\frac{\delta n_i + \delta n_j}{2 n_0} -
\sqrt{(1+\delta n_i/n_0)(1+\delta n_j/n_0)}\cos(\varphi_i - \varphi_j)
\right]
\right\} d \tau,
\nonumber
\\
\delta n_i \equiv n_i - n_0 = \delta n_i(\tau),\qquad
\varphi_i = \varphi_i(\tau).
\nonumber
\end{eqnarray}

From Equation (\ref{Z_2}) one can see, that increasing of the
mean number of particles
at each granule provided that $J = t n_0$ and $U$ are constant,
enables one to leave an action in terms of the phase degrees of
freedom alone
\cite{Bruder,Fisher} to lead to the action of quantum XY model.

Being interested in the difference between phase boundaries of models
(\ref{hamiltonian}) and (\ref{hamiltonian_xy}) at sufficiently high but
finite $n_0$, let us expand the superfluid density in powers of $1/n_0$
up to a second order. Defining the superfluid density from the response
of a system to the shift of phases at the boundary \cite{Barber},
from Equation (\ref{Z_2}) we have:
\begin{eqnarray}
\nu_{\bf s} = \gamma + \frac{1}{2n_0^2} \Gamma^{(2)} + ...
\label{series}
\end{eqnarray}
where $\gamma$ is the helicity modulus of quantum XY model
\cite{Minhagen,Jose}.
The first order corrections are equal to zero due to the
invariance of the action of the XY model against the "time" inversion.

Rather a complex expression for  $\Gamma^{(2)}$
can be represented as some equilibrium value of
quantum XY model \cite{JETP}, which can be easily estimated {\it via.}
quantum Monte Carlo technique or different self - consistent apprximations
\cite{Akopov}.

Given a value of the coefficient $\Gamma^{(2)}$ of the expansion
(\ref{series})
as a function of control parameters $\Gamma^{(2)} = \Gamma^{(2)}(q,T)$,
one can construct an upper estimation of the phase boundary
$T^c_{\bf h}(q; n_0)$ of the boson Hubbard model (\ref{hamiltonian}):
\begin{eqnarray}
T^c_{\bf h}(q; n_0) \le T^c_{\bf xy}(q) \left\{ 1 +
\frac{\nu_{\bf s}(q,T^c_{\bf xy}) - \gamma(q,T^c_{\bf xy})}
{\gamma(q,T^c_{\bf xy})} \right\} =
T^c_{\bf xy}(q) \left\{ 1 +
\frac{\pi \Gamma^{(2)}(q,T^c_{\bf xy})}{4 n_0^2 T^c_{\bf xy}(q)} \right\}
\label{correct_funcint}
\end{eqnarray}
where $T^c_{\bf xy}(q)$ is the line of topological phase transitions
of quantum XY model (\ref{hamiltonian_xy}).
The estimation (\ref{correct_funcint}) can be easily obtained from the
assumption that
lines of phase transitions of both models are defined by the
universal relation \cite{Minhagen}:
\[
\gamma(q, T^c_{\bf xy}) = 2 T^c_{\bf xy} / \pi, \;\;\;\;
\nu_s(q, T^c_{\bf h}) = 2 T^c_{\bf h} / \pi
\]
Results of above mentioned estimations are given in Figure~1.
It turns out that in the region $0.7 < q < 1.5$
the line of phase transitions of the Hubbard model approaches
(to within $5\%$) its limit at
$n_0 = 8$, whereas some greater densities $n_0 > 16$
are required in the strong quantum region $q > 1.7$ because of
the rapid increase of the coefficient $\Gamma^{(2)}(q)$.
Direct Monte Carlo calculation of the phase diagram $T^c_{\bf h}(q)$
(see below), being in agreement with the predictions of
$1/n_0$ - expansion at $0.7 < q < 1.5$, shows that in the case of
{\it strongly} interacting system (at $q > 1.7$)
theoretical estimations markedly overestimate the maximum boson
density $n_0$ at which mesoscopic effects are still essential.

To conclude of this chapter one additional feature should be recognized.
It is easy to see, that all estimations presented in this Section
and started from the partition function (\ref{Z_1})
have been carried out in the {\it grand canonical} ensemble with
zero chemical
potential. This has enabled us to make all calculations analytically,
disregarding the restrictions on the total number of particles in the
system
when integrating over the fluctuations of moduli of the order parameter.
In order to justify the possibility of comparing the MC results with
theoretical estimations it is need to show that,
being calculated within the same approach, the discrepancy $\delta n$
between the mean number of bosons per granule and $n_0$ is small.

From (\ref{Z_2}) for $\delta n$ one can write
\begin{eqnarray}
\delta n = \frac{1}{n_0} \Delta^{(1)} + ...
\label{series1}
\end{eqnarray}
As calculation shows, the value of $\Delta^{(1)}$ is less than $0.1$
in the region $q>0.7,\; T<1$. This observation justifies our use of
great canonical ensemble with zero chemical potential in estimating
the coefficient $\Gamma^{(2)}$ of series (\ref{series}).

\section{Quantum Monte Carlo simulations.}
\label{MC}

In studies of properties of the boson Hubbard model (\ref{hamiltonian})
at different values of control parameters $\{q,T\}$ we have applied
the "checkerboard" breakup \cite{Blaer}. In this method
the classical degrees
of freedom to be sampled are imaginary - timedependent boson
occupation number field $\{n_{i}^p\}$, $i = 1 \ldots N^2,\; p = 0 \ldots 4P$.
The algorithm of Monte Carlo (MC) calculations of
the quantum XY model was described in \cite{Jose,Belous}.
We have performed extensive tests to verify that our results converge
in the limit of $P \to \infty$. Results presented below have been
obtained by the averaging over $3 - 5$ initial configurations formed by
$4P$ - times ($P$ - times for the XY model) multiplication of the
random configuration of bosons (phases for the XY model) at the
lattice $N \times N$.

The main attention has been given to the calculation of the superfluid
density $\nu_s$. This quantity have been determined from
the winding number  \cite{Trivedi,Blaer}
\begin{eqnarray}
\label{nu_s_1}
\nu_{\bf s} = 0.5 T \left< W_{\bf x}^2+W_{\bf y}^2 \right>_{\bf h} \\
W_{\bf x} = \sum\limits_{p=0}^{4P}
\sum\limits_{i_y=1}^{N} (-1)^{i_x+p}n_{i}^{p},\;\;\;
W_{\bf y} = \sum\limits_{p=0}^{4P}
\sum\limits_{i_x=1}^{N} (-1)^{i_y+p}n_{i}^{p},
\nonumber
\end{eqnarray}
where $n_{i}^{p}$ means the number of bosons at a site $i$
(with coordinates $\{i_x,i_y\}$) of a level $p$ of 3D classical system.
We have also used the current autocorrelation function \cite{Tobochnik}
\begin{eqnarray}
\label{nu_s_2}
\nu_{\bf s} = - \frac{1}{n_0 N^2}
\left< \hat T_{\bf x} \right>_{\bf h} - \frac{1}{n_0^2 N^2 T P}
\sum\limits_{\tau=0}^{P-1} \left< \hat J_{\bf x}^{(p)}(\tau)
\hat J_{\bf x}^{(p)}(0) \right>_{\bf h}
\\
\hat T_{\bf x} = -\frac{1}{2} \sum\limits_i
\left\{ a{^\dagger}_{i+x} a_i +
       a{^\dagger}_i a_{i+x} \right\},
\nonumber
\\
\hat J_{\bf x}^{(p)} = - \frac{\tilde \jmath}{2} \sum\limits_i
\left\{ a{^\dagger}_{i+x}a_{i} -
       a{^\dagger}_{i}a_{i+x} \right\},\;\;\;
\hat J_{\bf x}^{(p)}(\tau) = e^{\tau \beta \hat H / P} \hat J_{\bf x}^{(p)}
e^{-\tau \beta \hat H / P}
\nonumber
\end{eqnarray}
The substitution $a_i \to \sqrt{n_0} e^{\i \varphi_i}$
transforms Equation (\ref{nu_s_2}) to the well - known expression for the
helicity modulus $\gamma$ of the quantum XY model \cite{Jose}.

As have been pointed out by Scalapino et. al. \cite{Scalapino},
the temperature derivative of the superfluid density
gives an additional information about the type of phase transition
at some temperature $T^c(q)$: in framework of the Kosterlitz - Thouless
picture the value of $\partial ( \beta \nu_{\bf s} ) / \partial \beta$
scales to a Dirac delta function $\delta(T-T^c)$.
On the finite lattices $\partial ( \beta \nu_{\bf s} ) / \partial \beta$
shows a response which increases with lattice size $N$, the
position of the maximum of the derivative being independent
on $N$.

To find the derivative of the superfluid density
$\partial ( \beta \nu_{\bf s} ) / \partial \beta$ we have estimated
the difference in internal energies of systems which differ
by a phase twist $\delta \varphi$ in the boundary condition along one
lattice direction.
\begin{eqnarray}
\frac{\beta\{ E(\delta \varphi) - E(0) \}}{n_0 \beta t} \sim
\nu_{\bf s} + \beta \frac{\partial \nu_{\bf s}}{\partial \beta}
\label{dE}
\end{eqnarray}
One can show, that for the Cooper pairs of charge $2e$
this phase twist can be realized in the "flux quantization" scheme
and is equivalent to threading a flux through the center of a thorus on
which the system lies \cite{Scalapino}.

We have also calculated  he fluctuation of bosons at lattice sites
\begin{eqnarray}
\delta n^2_{\bf h} = \frac{1}{4P N^2}
\left< \sum\limits_{p=0}^{4P-1}\sum\limits_{i}
\{n_{i}^p - n_0\}^2 \right>_{\bf h}
\label{n_fluct}
\end{eqnarray}

\section{Results and discussion.}
\label{discuss}

Shown at Figure~2 are dependencies of the superfluid fraction
$\nu_{\bf s}(T)$ of the Hubbard model at $q = 0.2$
(in the classical region of XY model (\ref{hamiltonian_xy}), Figure~2a)
and $q = 2.0$ (see Figure~2b). For reference, the helicity modulus
$\gamma$ of quantum XY model as function of temperature $T$ is also
plotted. Analysis of data obtained at
different sizes $N$ and densities $n_0$ of the system reveals that for
the system of strongly interacting bosons (at $q=2.0$) the MC results are
in qualitative agreement with the theoretical estimations of Sections
\ref{theor_mfa} and \ref{theor_exp}.
Really, from Figures 1 and 2 one can see that as the density of bosons
$n_0$ is increased, the boundary of
ordered superconducting state of the system (\ref{hamiltonian})
approaches that of quantum XY model with critical temperatures $T_{\bf h}^c$
of the Hubbard model being greater than $T_{\bf xy}^c$ of quantum XY model.
The line of transitions $T^c(q)$ can be estimated from the universal
relation $\nu_{\bf s}(T^c) = 2T^c / \pi$. Thus defined, the temperature of
metal - superconductor transition agree fairly well with the position
of the peak of the temperature derivative of superfluid density (\ref{dE}).
Our calculation shows, that the position of the maximum of the derivative
does not depend (to within statistical errors) on the system size
and, as one can see from Figure~3, lowers as $n_0$ is increased.

From Figure~2a one can show, that the transition temperature of the
weakly interacting bosons (at $q = 0.2$) appeared to be {\it less}
than that of quantum XY model. This tendency persists with increasing
the system size.
The theoretical approach  used in Section \ref{theor_exp}
can not help to elucidate the reason of this phenomena because
at $q < 0.4$ relative fluctuations of moduli of the
order parameter are only weakly damped by the interaction,
corrections $\Gamma^{(2)}$ and $\Delta^{(1)}$ are large and theoretical
estimations work badly.

Let us consider the behaviour of $\nu_s(q)$ and $\gamma(q)$
as functions of quantum parameter $q$ at $T = 0.5$ (see Figure~4).
Defining the point $q^c$ of phase transition from the universal
relation $\nu_{\bf s}(q) = 2 T/\pi$ we see
that the boundary of ordered superconducting state of the model
(\ref{hamiltonian}) lies to the right of the boundary of the model
(\ref{hamiltonian_xy}).
This conclusion is verified by the results of calculations
of the derivative $\partial ( \beta \nu_{\bf s} ) / \partial \beta$
presented at the insert of Figure~4.
The position $q|_{n_0=3} \approx 2.3$ of a peak of the
derivative is in a sufficiently good agreement with the critical
point $q^c|_{n_0=3} \approx 2.4$ determined from the universal
relation.

The dependence of the relative fluctuation of particle number
at sites of the system {\it vs.} quantum parameter $q$ is shown at Figure~5.
Particularly, Figure~5 can serve as an illustration of the
role of interaction in the transition to the quasiclassical limit
from the boson Hubbard to quantum XY model. Really, at finite
densities $n_0$ the spectrum of the operator $\hat n_i - n_0$
can be considered as unbounded only if relative fluctuations of
particle number are small $\delta n^2_{\bf h} / n_0^2 \ll 1$.
Then, as it is usually done on examination Josephson or granular systems
in terms of the model (\ref{hamiltonian_xy}),
the particle number operator $\hat n_i - n_0$
can be chosen as a conjugate one to the "phase" operator $\hat \varphi_i$:
$\hat n_i - n_0 = \i \partial / \partial \varphi_i$
Increasing of the interaction (quantum parameter $q$) leads to the
suppressing of the relative fluctuations of the order parameter module
as can be seen from Figure~5. It should be noted, that at high $q$ the
fluctuations of particles number are {\it greater} than those of
quantum XY model and approach their with increase in density.
The results presented in the insert of the Figure are
relative fluctuations $\delta n^2_{\bf h} / n_0^2$ as functions of
quantum parameter $q$ at $T = 0.5$ at different densities $n_0$.
The increase in $n_0$ is seen to be of great importance in suppressing
the relative fluctuations.

A great attention has been given thus far to the possibility of
{\it reentrance} phenomena, when the global superconducting state
in some region of $q$  is absent not only at high, but also at
sufficiently low temperatures.
In the framework of XY model the possibility of this phenomenon taking place
has been connected with the domain of phases \cite{Simanek}, dissipation or
mutual capacitances effects \cite{Choi,Stroud}. From results presented
one can see that taking into account the fluctuation of moduli of the order
parameter does not lead to the reentrance phenomena at least in the region
explored.

To conclude, we have used the boson Hubbard model to analyze
the effect of quantum fluctuations of
phases and moduli of the order parameter on the onset of superconductivity
in 2D mesoscopic Josephson system.  Both mean field approximation
and $1/n_0$ expansion leads to the conclusion that the line $T^c(q)$ of
superconductor - metal phase transitions  lies {\it above}
that of quantum XY model, the latter being quasiclassical limit
(as $n_0 \to \infty, \; U \ne 0$) of the Hubbard one.
Our MC simulations show that in the region $q < 1$
of small quantum fluctuations of phases it needs an average of $10$
bosons per site to suppress relative fluctuations of the local superfluid
density. As interaction is increased, the quasiclassical
limit is approached at lower densities ($n_0 \sim 8$ at $q \sim 2$).
No reentrance or discontinuity phenomena have been found.

This work was supported by Russian Foundation of Basic Research, Programs
"Solid state Nanostructures" and Superconductivity.

\newpage
Figure~1. \\
The phase diagram of 2D Hubbard (\ref{hamiltonian}) and quantum XY
(\ref{hamiltonian_xy}) models. S - supercondudting, N - normal state.
The mean - field results: {\bf 1}: $n_0=1$, {\bf 2}: $n_0=2$,
{\bf 3}: $n_0=6$, {\bf 4}: quantum XY model ($n_0=\infty$);
The $1/n_0$ expansion (\ref{correct_funcint})
results: {\bf 5}: $n_0=6$, {\bf 6}: $n_0=14$.
Symbols present points of phase transitions have been found by MC method.

\vspace{0.5 cm}

Figure~2a. \\
The superfluid density $\nu_{\bf s}$ (helicity modulus $\gamma$) {\it vs.}
temperature $T$ at $q=0.2$.
The dependence $2T / \pi$ (see in text) is given by a dashed line.
Data are connected to guide the eyes.
If not presented, error bars are within the size of the data point.
\vspace{0.5 cm}

Figure~2b. \\
The superfluid density $\nu_{\bf s}$ (helicity modulus $\gamma$) {\it vs.}
temperature $T$ at $q=2.0$.
\vspace{0.5 cm}

Figure~3. \\
$T \beta \{ E(\pi/2) - E(0) \} / \{n_0 N^2\}$
as a function of temperature $T$.
The solid lines are spline fits to the data.
\vspace{0.5 cm}

Figure~4. \\
The superfluid density $\nu_{\bf s}$ (helicity modulus $\gamma$) {\it vs.}
quantum parameter $q$ at $T=0.5$. The line $1 / \pi$
is shown with the help of a dashed line. \\
Insert: the temperature derivative of the superfluid fraction (\ref{dE}).
\vspace{0.5 cm}

Figure~5. \\
Fluctuation $\delta n^2_{\bf h}$ of the number of bosons
at sites of the system
as a function of quantum parameter $q$ at $T = 0.5$. \\
Insert: relative fluctuations $\delta n^2_{\bf h} / n_0^2$ at
different densities $n_0$.

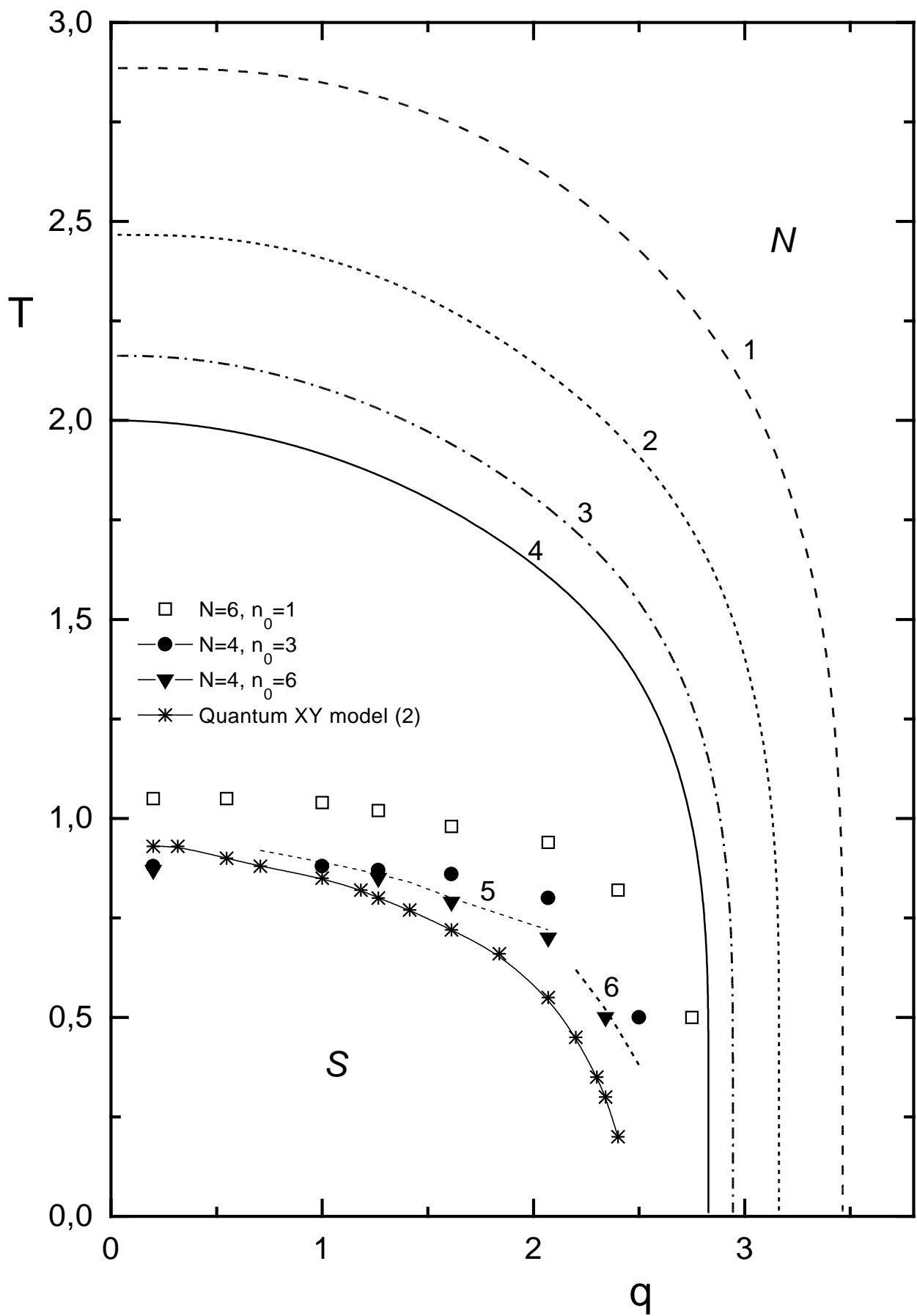
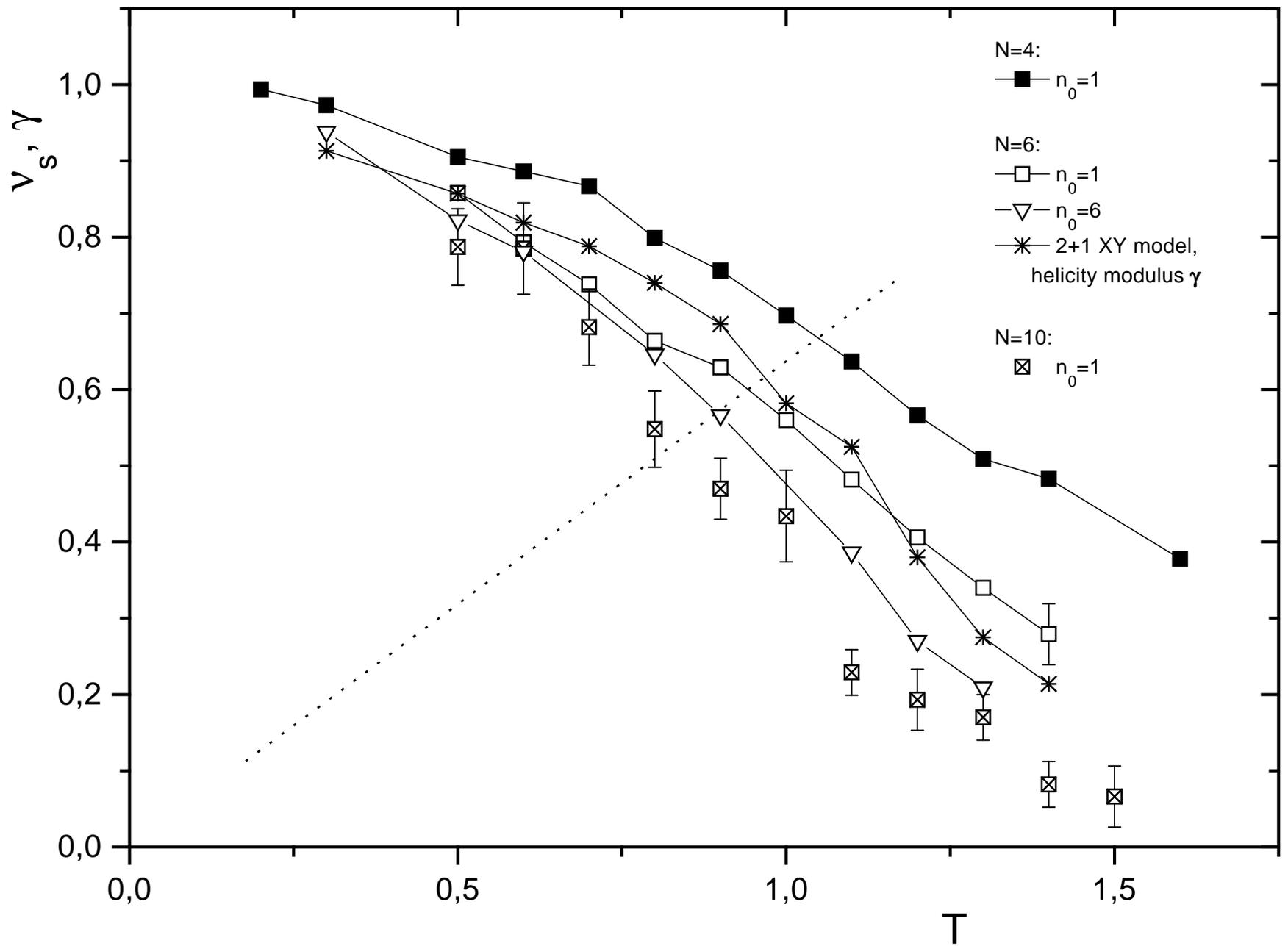
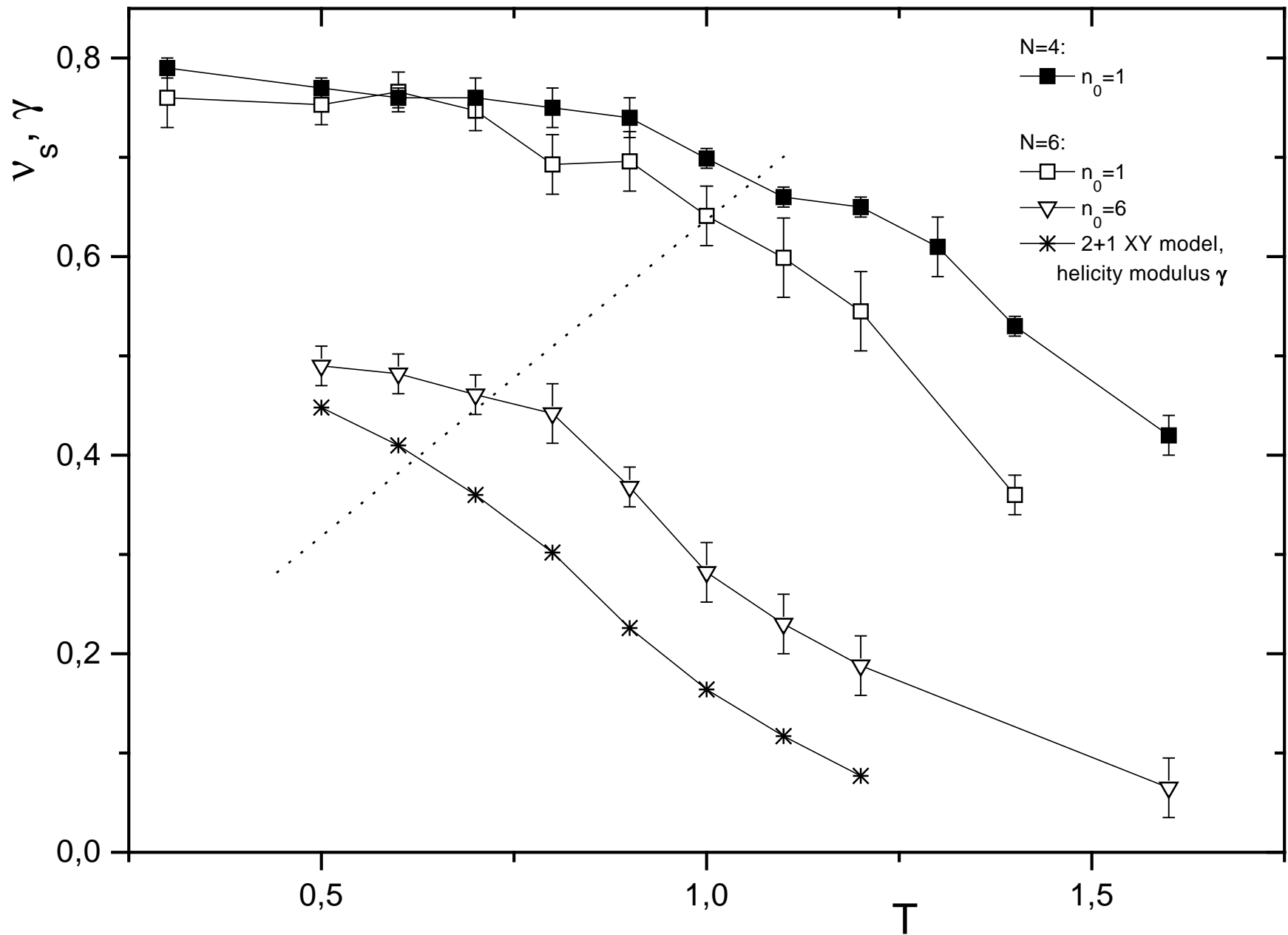
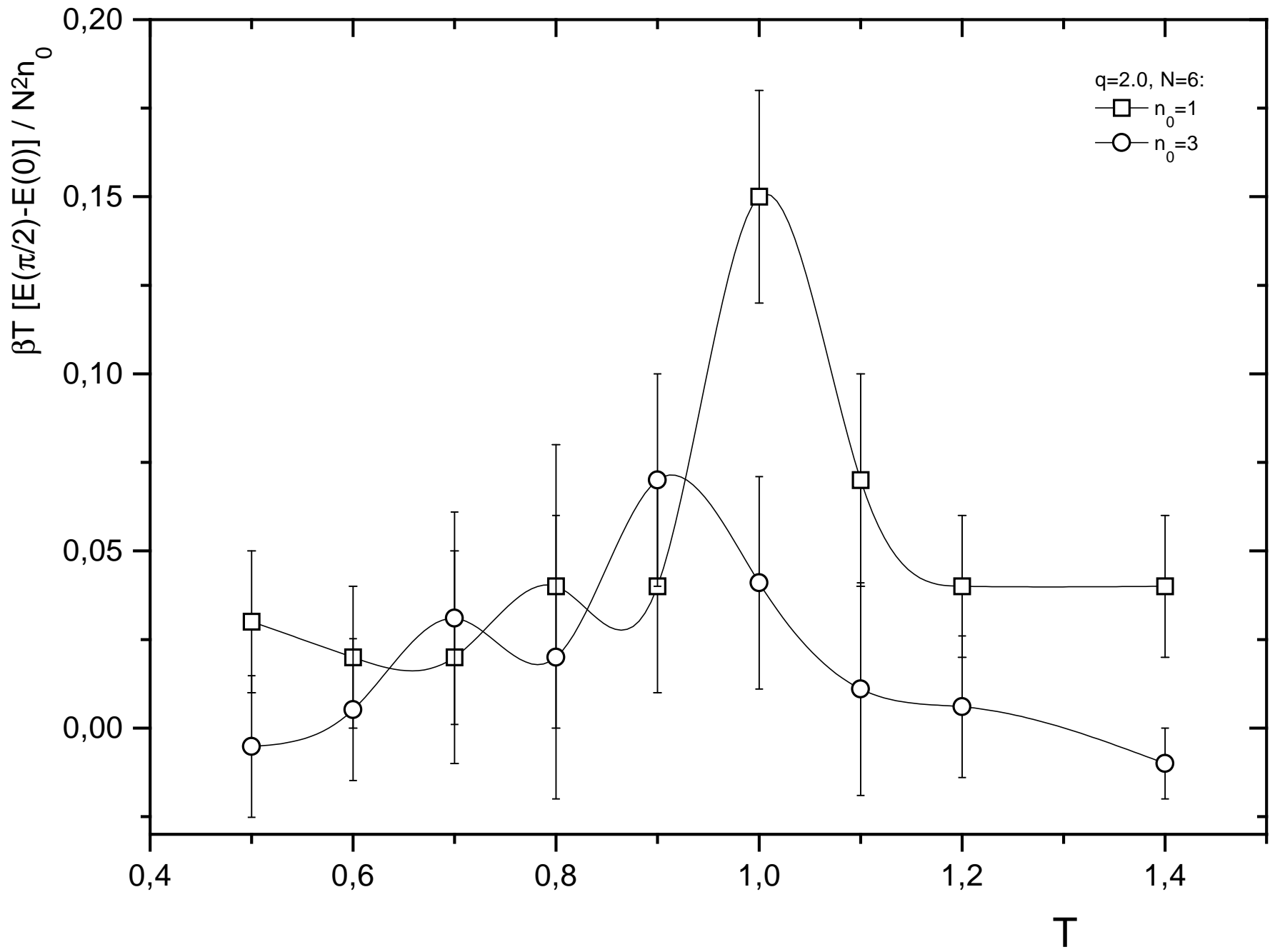
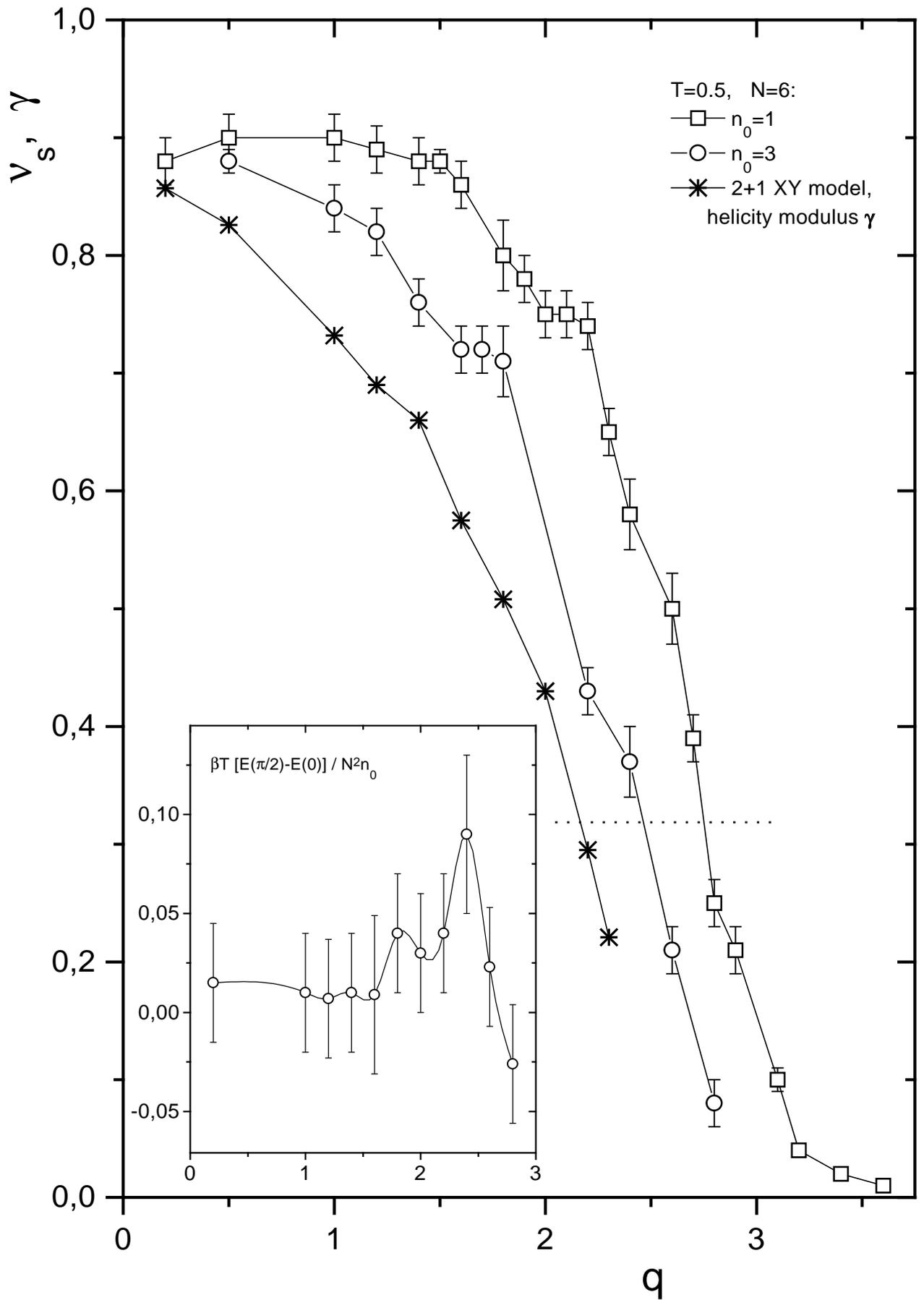
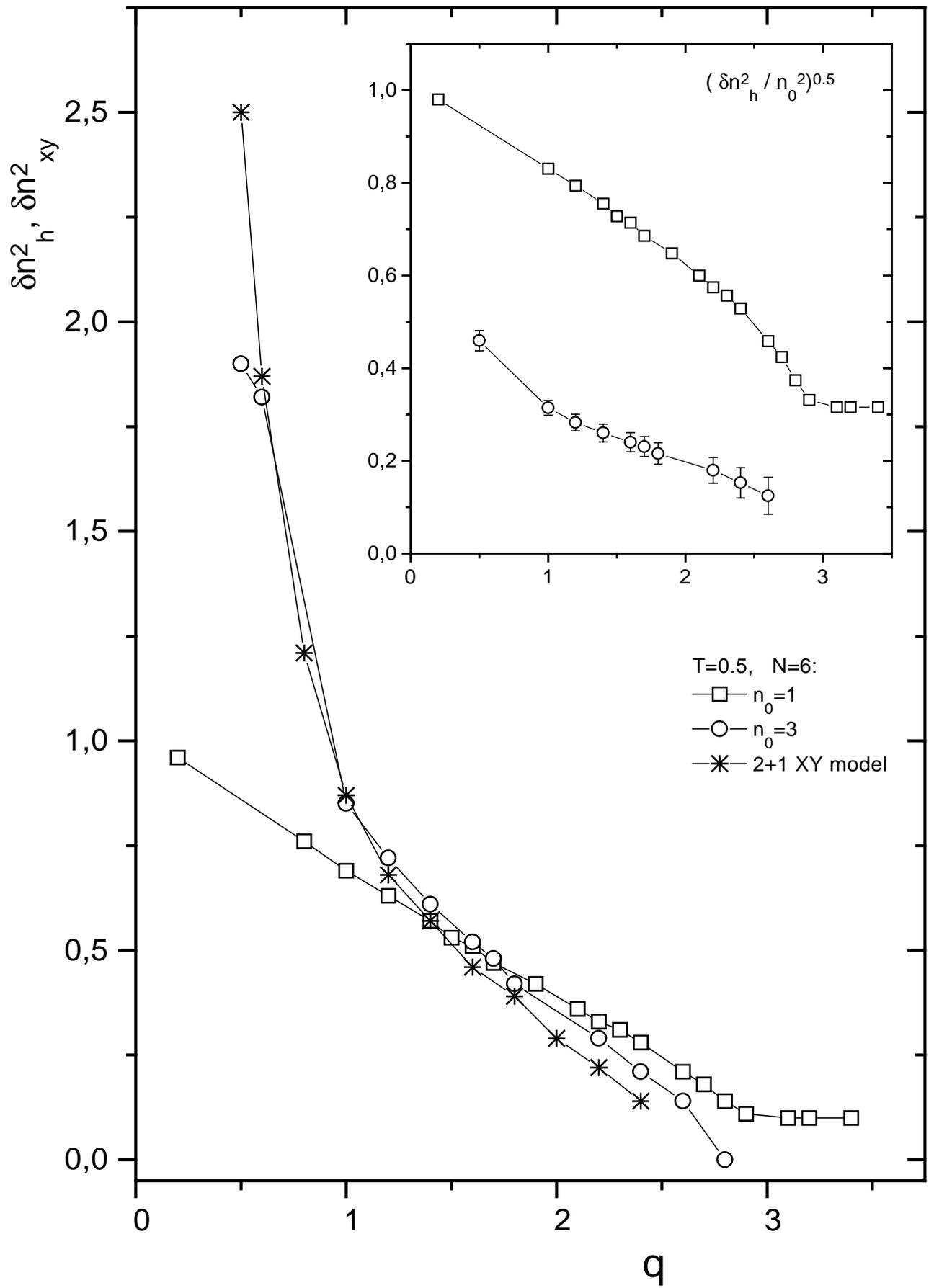
\end{document}